\documentstyle[prb,aps]{revtex}
\begin{document}
\draft
\title{Magnetic excitations and structural change in\\
the $S$=$\frac{1}{2}$ quasi-one-dimensional magnet\\
Sr$_{14-x}$Y$_x$Cu$_{24}$O$_{41}$ (0$\leq x \leq$1)}
\author{M. Matsuda and K. Katsumata}
\address{
The Institute of Physical and Chemical Research (RIKEN),
Wako, Saitama 351-01, Japan}
\author{T. Osafune, N. Motoyama, H. Eisaki, and S. Uchida}
\address{
Department of Superconductivity, The University of Tokyo,
Bunkyo-ku, Tokyo 113, Japan}
\author{T. Yokoo*, S. M. Shapiro, and G. Shirane}
\address{
Department of Physics, Brookhaven National Laboratory,
Upton, New York 11973}
\author{J. L. Zarestky}
\address{
Ames Laboratory, Ames, Iowa 50011}
\date{Received 12 June 1997}
\maketitle
\begin{abstract}
Neutron scattering measurements have been performed
on the $S$=$\frac{1}{2}$ quasi-one-dimensional system
Sr$_{14-x}$Y$_x$Cu$_{24}$O$_{41}$, which has both
simple chains and two-leg ladders of copper ions. We observed
that when a small amount of yttrium is substituted for strontium,
which is expected to
reduce the number of holes, the dimerized state and the structure 
in the chain are changed drastically. The inelastic peaks originating
from the dimerized state of the chain becomes broader in energy
but not in momentum space. This implies that the dimerized
state becomes unstable but the spin correlations are unchanged
with yttrium substitution. Furthermore, it was observed that nuclear
Bragg peak intensities originating from the chain show strong
temperature and $x$ dependence, which suggests that the chains
slide along the $c$ axis as temperature and $x$ are varied.
\end{abstract}

\section{Introduction}
As a byproduct of the high temperature superconducting copper
oxides, many interesting quasi-one-dimensional copper oxides have
been discovered or rediscovered recently.
Sr$_{14}$Cu$_{24}$O$_{41}$ is one of them and consists of two
kinds of unique building blocks as shown in Fig. 1a.
\cite{ta10,ta13} One is simple chains of copper ions which are
coupled by the nearly 90$^\circ$ Cu-O-Cu bonds. The other is
two-leg ladders of copper ions, which are coupled by the nearly
180$^\circ$ Cu-O-Cu bonds along the $a$ and $c$ axes. This
compound has been extensively studied since both of the building
blocks show interesting ground states.

The ground state of a two-leg spin ladder system
is a singlet state as observed in SrCu$_2$O$_3$. \cite{ta16}
It was also shown that the ladder in the related compound,
Sr$_{14}$Cu$_{24}$O$_{41}$ has a singlet ground state
with a fairly large gap. \cite{ecc,kuma,kita}
The property of the energy gap in spin ladders is interesting from
the view point of quantum phenomena in a low-dimensional
(between 1 and 2)
Heisenberg antiferromagnet. It was theoretically predicted that the
spin $\frac{1}{2}$ Heisenberg ladder with even numbers of legs
has an excitation gap and that the excitation is gapless for the spin
ladder with odd numbers of legs. \cite{ta1}
The two-leg ladder system has
also attracted many researchers since superconductivity is expected
in the carrier doped spin ladder system. \cite{ta2,ta3} Recently,
Uehara $et$ $al.$ \cite{aki} found that
Sr$_{0.4}$Ca$_{13.6}$Cu$_{24}$O$_{41}$ shows
superconductivity below 12 K under a high pressure of 3 GPa.
The superconductivity in the ladder system is considered to be
crucial to understand the mechanism of the high
temperature superconductivity.

As Matsuda $et$ $al.$ \cite{matsu0,matsu1} showed, the simple
chain in Sr$_{14}$Cu$_{24}$O$_{41}$ also has an interesting
singlet ground state originating from a dimerization. Surprisingly,
the dimers are formed between
spins which are separated by 2 and 4 times the distance between
the nearest-neighbor (n.n.) copper ions in the chain.
This is probably related to localized holes in the
chain which are expected to make the interaction between copper
spins longer-ranged. The trivalent yttrium substitution for divalent
strontium is expected to decrease hole carriers.
The dimerized state in the chain depends critically on the number
of holes ($N_h$). The magnetic inelastic peaks become broader
in energy with yttrium substitution. \cite{matsu1} Furthermore,
La$_6$Ca$_8$Cu$_{24}$O$_{41}$ in
which $N_h$=0 shows a long-range magnetic order with
ferromagnetic correlations within the chain. \cite{carter,matsu2}

We have performed neutron scattering experiments
to study the effects of yttrium substitution on the magnetic and
structural properties of the chains in
Sr$_{14}$Cu$_{24}$O$_{41}$.
We previously reported the results of neutron scattering
experiments using polycrystalline samples of
Sr$_{14-x}$Y$_x$Cu$_{24}$O$_{41}$ ($x$=0, 1, and 3).
\cite{matsu1} In order to make a detailed study of the magnetic
excitation in a wide range of energy ($\omega$) - momentum
($Q$) space and the crystallographic structure, we performed
new experiments on high quality single crystals. Furthermore,
we concentrated on the samples with low concentration of yttrium
to study systematically how the dimerized state is changed since
the dimerized state is destroyed with only a small amount of
yttrium substitution. \cite{matsu1} It was
observed that when yttrium is lightly substituted for strontium,
strong and sharp magnetic inelastic peaks which originate from the
dimerized state in the chain become broader. With further yttrium
substitution, the inelastic peaks become much broader and the
excitation energy is decreased. The interesting point
is that the inelastic peaks become broader only in energy but not
in momentum space. This means that the dimerized state becomes
unstable but the spin correlations are unchanged with yttrium
substitution. It was also observed that nuclear Bragg intensities
originating from the chain show strong temperature and yttrium
concentration dependence. One possible explanation for this
would be that the chains shift along the $c$ axis with temperature
and yttrium substitution.

The format of this paper is as follows: Experimental details are
described in Sec. II. The magnetic and structural studies are
presented in Sec. III and IV, respectively. In Sec. V
the experimental results are discussed.

\section{Experimental Details}
The single crystals of Sr$_{14-x}$Y$_x$Cu$_{24}$O$_{41}$
($x$=0, 0.10, 0.25, and 1.0) were grown using a traveling solvent
floating zone (TSFZ) method at 3 bars oxygen atmosphere. The
effective mosaic of the single crystals is less than 0.4$^\circ$ with
the spectrometer conditions as is described below. It is expected that
yttrium is distributed homogeneously in the sample since the lattice
constant $b$ is systematically decreased
and the linewidth of the nuclear Bragg peaks does not change with
yttrium substitution. The lattice constant $b$ is 13.36 $\AA$ and
13.08 $\AA$ at 10 K for the $x$=0 and $x$=1 samples,
respectively.

The neutron scattering experiments were carried out on
the HB3 triple-axis spectrometer at the High Flux Isotope Reactor
at Oak Ridge National Laboratory and
the H8 triple-axis spectrometer at the High Flux Beam
Reactor at Brookhaven National Laboratory. The horizontal
collimator sequence was open-40'-S-60'-120' for the experiments
on the HB3 and 40'-40'-S-80'-80' on the H8. The neutron
measurements of Sr$_{14}$Cu$_{24}$O$_{41}$,
Sr$_{13.9}$Y$_{0.1}$Cu$_{24}$O$_{41}$, and
Sr$_{13.75}$Y$_{0.25}$Cu$_{24}$O$_{41}$ were performed
on HB3 and Sr$_{13}$Y$_{1}$Cu$_{24}$O$_{41}$ on
H8. The final neutron energy
was fixed at $E_f$=14.7 meV. Pyrolytic graphite crystals
were used as monochromator and analyzer; contamination from
higher-order beam was effectively eliminated using a pyrolytic
graphite filter after the sample. The single crystals were
mounted in closed-cycle refrigerators which allowed us to
perform the measurements over a wide temperature range
10 - 300 K. The experiments for scattering in the
$(0,k,l)$ zone were performed. As described in Ref.
\onlinecite{ta10},
there are three different values for the lattice constant $c$.
Since we will mainly show the magnetic and structural properties
in the chain, $c_{chain}$ will be used to express Miller
indices.

\section{Magnetic Excitations}
Figure 1b shows the temperature dependence of magnetic
susceptibility parallel to the $c$ axis in single crystals of
Sr$_{14-x}$Y$_x$Cu$_{24}$O$_{41}$ ($x$=0, 0.25, and 1.0).
Since the spin gap originating from the ladder has a large value of
$\sim$400 K, \cite{ecc,kuma,kita} the susceptibility below room
temperature comes predominantly from the chain. The
susceptibility in
Sr$_{14}$Cu$_{24}$O$_{41}$ shows a broad peak around
80 K and the Curie-Weiss tail can be seen at low temperatures. The
Curie-Weiss term is increased and the broad peak is shifted to
lower temperature with yttrium substitution.

We show in Fig. 2 inelastic neutron scattering spectra at $T$=10
K observed at (0,3,0.085), (0,3,0.17), and (0,3,0.25) in single
crystals of Sr$_{14-x}$Y$_x$Cu$_{24}$O$_{41}$ ($x$=0 and
0.25). Note that the indices correspond to (0,3,0.12)$_{ladder}$,
(0,3,0.24)$_{ladder}$, and (0,3,0.36)$_{ladder}$, respectively.
\cite{matsu1} Two sharp, intense inelastic peaks are observed in
Sr$_{14}$Cu$_{24}$O$_{41}$. The peaks are the sharpest at
(0,3,0.17) since the resolution ellipsoid is almost parallel to the
dispersion curve so that the focusing effect is expected. Thus, the
intrinsic linewidth of the inelastic peaks we observed here is
almost resolution-limited. Note that in the previous paper
\cite{matsu1} we reported the constant-$Q$ scans at (0,3,-$L$)
since the focusing condition was different for the spectrometer.
The inelastic peak positions slightly change with $Q$ which
follows the $\omega$-$Q$ dispersion relation \cite{matsu1} as
shown in the inset of Fig. 2.
One puzzling feature is the presence of two excitations
originating from the chain. The presence of two
peaks could be due to the anisotropy in fluctuations parallel
and perpendicular to the chain direction or the presence of
other interactions.

In Sr$_{13.75}$Y$_{0.25}$Cu$_{24}$O$_{41}$ the linewidth
of the inelastic peaks becomes broader. Whereas, the peak positions
are almost unchanged. These results indicate that the dimerized
state becomes unstable with yttrium substitution.
As described above, the broad peak in susceptibility is shifted to
lower temperature in
Sr$_{13.75}$Y$_{0.25}$Cu$_{24}$O$_{41}$. This is probably
because an increase of the Curie-Weiss tail shifts the peak
to lower temperature even though the intrinsic peak position is
almost unchanged.

Figure 3a shows a constant-$Q$ scan at $T$=10 K observed at
(0,3,-0.14) in Sr$_{13}$Y$_1$Cu$_{24}$O$_{41}$. The focusing
effect is also expected at (0,3,-0.14) as at (0,3,0.17). The inelastic
peaks become much broader and spread over from 6 to 13 meV.
This behavior is consistent with that observed in the powder
sample. \cite{matsu1} Since we used a single crystal, it is also
possible to measure $\vec{Q}$-dependence of the magnetic
excitations. We show in Fig. 3b a constant-$E$ scan at
$\Delta E$= 8 meV observed at (0,3,-$L$). A broad peak can be
seen around $L_{chain}$=0.25. The peak position is similar to the
position at which the correlation function $S(Q)$ shows
a maximum in Sr$_{14}$Cu$_{24}$O$_{41}$, \cite{matsu1}
suggesting that spin correlations are unchanged with yttrium
substitution.

Figure 4 shows inelastic neutron scattering spectra of
Sr$_{14-x}$Y$_x$Cu$_{24}$O$_{41}$ ($x$=0 and 0.25)
at $T$=10 K observed at (0,0,1.1)$_{ladder}$.
In Sr$_{14}$Cu$_{24}$O$_{41}$ a resolution-limited sharp
peak in energy was observed around 12 meV.
Although the index at which the dispersion curve has a minimum
corresponds to the ladder, this peak does not originate from the
intra-ladder coupling since the spin gap energy is about 35 meV
\cite{ecc} and the dispersion curve is expected to have minima at
(0,0,$L_{ladder}$) when $L_{ladder}$=$n$+1/2 or at
($H_{ladder}$,0,0) when $H_{ladder}$=$n$+1/2 ($n$: integer).
In the previous paper \cite{matsu1}
we speculated that the peak originates from the dimerized state in
the ladder which is formed between the nearest-neighbor copper
ions connected by the inter-ladder coupling, although the number
of the dimers are considered to be small. Very recently Mikeska
and Neugebauer \cite{mike} showed that the spin gap due to the
inter-ladder coupling should be much larger than 12 meV. Then the
dimerization originating from the inter-ladder coupling would be
realized if a local distortion occurs due to the localized holes.
They also showed that the excitation at (0,0,1)$_{ladder}$ is
explained with the theory of non-magnetic impurities in decoupled
ladder. The sharp (0,0,1)$_{ladder}$ peak in
Sr$_{14}$Cu$_{24}$O$_{41}$ becomes much broader in
Sr$_{13.75}$Y$_{0.25}$Cu$_{24}$O$_{41}$. This strongly
suggests that the excitation is closely related with the hole in the
ladder, which probably couples with copper spin to form a singlet
as in the chain and acts as a non-magnetic impurity.

\section{Structural Change in the Chain}
Figure 5 shows the unusual temperature dependence of the nuclear
Bragg peak intensity originating from the chain for different values
of $x$. For $x$=0 (Fig. 5a) the peak intensity remains constant
below 30 K and then decreases with increasing temperature above
30 K. A surprising behavior was observed upon yttrium
substitution. In Sr$_{13.9}$Y$_{0.1}$Cu$_{24}$O$_{41}$
(Fig. 5b) the intensity decreases up to 60 K and increases above
60 K with increasing temperature. In
Sr$_{13.75}$Y$_{0.25}$Cu$_{24}$O$_{41}$ (Fig. 5c) the
intensity remains constant below 30 K and then increases above
30 K. Other nuclear Bragg peaks from the chain also show fairly
large temperature and yttrium concentration dependence. In
Sr$_{13.9}$Y$_{0.1}$Cu$_{24}$O$_{41}$, for
example, the intensity of the (0,2,2) Bragg peak shows similar
temperature dependence as that of (0,0,2).
The intensity at (0,0,4) decreases with increasing
temperature. On the other hand, the intensity at (0,1,1)
slightly decreases with increasing temperature.

Since ferromagnetic long-range ordering was observed in
La$_6$Ca$_8$Cu$_{24}$O$_{41}$ \cite{matsu2} which has no
holes, the (0,0,2) Bragg peak in
Sr$_{14-x}$Y$_{x}$Cu$_{24}$O$_{41}$ might originate from
a ferromagnetic ordering in the chain. The x-ray diffraction
experiments were performed in Sr$_{14}$Cu$_{24}$O$_{41}$
to clarify this point. \cite{cox} The results revealed that the
(0,0,2) Bragg peak intensity shows the same temperature
dependence as in Fig. 5a, indicating that the Bragg intensity is
nuclear in origin.

No major change of the intensity of the nuclear Bragg peak from
the ladder was observed for 10 $\leq T \leq$ 300 K.

\section{Discussion}
We observed a drastic change of magnetic and structural properties
with yttrium substitution. If the holes preferably exist on the chain
\cite{ta25}, one can estimate that the number of holes $N_h$ in the
chain is 60$\%$ of the copper ions in the chain in
Sr$_{14}$Cu$_{24}$O$_{41}$.
Whereas, in Sr$_{13.75}$Y$_{0.25}$Cu$_{24}$O$_{41}$
$N_h$ in the chain is estimated to be 57.5$\%$ of the
Cu ions in the chain. The change of $N_h$ is only
4.2$\%$. It is surprising that such a small change of $N_h$ would
affect the magnetic and structural properties so drastically.

Our motivation to perform the experiment using a sample with
low yttrium concentrations is to clarify how the intensity around
$L_{chain}$=1/8 and 1/4 changes with yttrium substitution. As
reported by Matsuda $et$ $al.$, \cite{matsu1} the dimers are
formed between copper spins which are separated by 2 and 4 times
the distance between n.n. copper ions in
Sr$_{14}$Cu$_{24}$O$_{41}$.
The trivalent yttrium substitution for divalent strontium is expected
to decrease hole carriers and thus increase the number of
Cu$^{2+}$ in the chain.
Then number of the dimers which are formed between copper spins
which are separated by 4 times the distance between n.n. copper
ions is expected to be decreased. With further yttrium substitution,
the dimers which are formed between n.n. copper spins are
expected to appear. Then we expect that the scattering around
$L_{chain}$=1/8 will first disappear and then the scattering
around $L_{chain}$=1/2 appear with increasing yttrium
concentration.

As shown in Fig. 2, the linewidth in energy becomes broader
with slight yttrium substitution. However, the integrated intensities
of the inelastic peaks over energy are just slightly decreased around 
$L_{chain}$=1/8 and 1/4. Furthermore, we did not observe
distinct inelastic peaks around $L_{chain}$=1/2 in
Sr$_{13.75}$Y$_{0.25}$Cu$_{24}$O$_{41}$ or
in Sr$_{13}$Y$_{1}$Cu$_{24}$O$_{41}$.
Thus, the inelastic peaks become broader only in energy but not
in momentum space, i.e. $S(Q)$ is unchanged. This implies that the
dimerized state becomes unstable but the spin correlations are
unchanged with yttrium substitution.
These results are consistent with the speculation on the dimerized
state in Ref. \onlinecite{matsu2} that the dimerized state in the
chain becomes unstable because the reduction of the holes makes
the ferromagnetic nearest-neighbor interactions more dominant and
the antiferromagnetic further-neighbor interaction less dominant.
This is deduced, in part, from the fact that
La$_6$Ca$_8$Cu$_{24}$O$_{41}$ shows a ferromagnetic
long-range order in the chain.
The remaining puzzle is why the scattering around
$L_{chain}$=1/8 is not suppressed with increasing yttrium
concentration.

We also observed that when yttrium is lightly substituted for
strontium ($x \leq$0.25), the gap energies are almost unchanged.
With further yttrium substitution ($x$=1.0), the excitation energy
is decreased. This suggests that the exchange interaction between
the spins which form the dimer is mediated by the hole at the
oxygen site and that the hole probably makes the interaction
longer-ranged due to the hopping mechanism. The longer-ranged
exchange interaction becomes weaker when $N_h$ is reduced.

We now discuss the structural change in the chain as observed by
the changes in nuclear Bragg peak intensities (Fig. 5). It is known
that the lanthanide or calcium substitution for strontium affects the
crystal structure. \cite{ta10,ta13} Adjacent chains are staggered
in Sr$_{8}$Y$_{6}$Cu$_{24}$O$_{41}$,
La$_{6}$Ca$_{8}$Cu$_{24}$O$_{41}$, and
Sr$_{8}$Ca$_{6}$Cu$_{24}$O$_{41}$ (Fig. 6a-i),
whereas the chains of Sr$_{14}$Cu$_{24}$O$_{41}$
slightly shift alternately along the $c$ axis (Fig. 6a-ii).
However, the structure of the ladder and strontium layers does not
change with the lanthanide or calcium substitution.
This suggests that the chains are almost independent of the
strontium and ladder layer, which form a solid structure, and
movable along the $c$ axis relatively.
In order to explain the temperature and yttrium substitution
dependence of the nuclear Bragg intensity from the chain we
performed a qualitative model calculation. Due to the small
number of the Bragg reflections purely from the chain,
it is difficult to determine the structure of the chain quantitatively.
In Fig. 6a we show simple models which describe the shift of
the copper ions.
We also assumed that both the copper and oxygen ions shift
along the $c$ axis synchronously.
The only parameter is a deviation $\delta$ along the $c$ axis.
$\delta$ equals to 0.33 at room temperature in
Sr$_{14}$Cu$_{24}$O$_{41}$. \cite{ta10}
Figure 6b shows the calculated intensity of (0,0,2) when $\delta$ is
varied from 0 (Fig. 6a-i) to 0.5 (Fig. 6a-iii).
In the calculation the contribution from the copper and oxygen
in the chain was considered. The calculated intensity has maxima at
$\delta$=0 and 0.5 and a minimum at $\delta$=0.25. We try to
explain the experimental results based on this model. Since the
scattering intensity of (0,0,2) observed in the $x$=0 sample increases
with decreasing temperature, $\delta$ should become larger
monotonically at lower temperature. To explain the temperature
dependence of the nuclear Bragg intensity in
Sr$_{13.9}$Y$_{0.1}$Cu$_{24}$O$_{41}$, $\delta$ should
be slightly below 0.25 at room temperature and increase with
decreasing temperature.
In Sr$_{13.75}$Y$_{0.25}$Cu$_{24}$O$_{41}$ $\delta$ should
be further decreased. This yttrium substitution dependence of
$\delta$ is consistent with the fact that adjacent chains in
Sr$_{8}$Y$_{6}$Cu$_{24}$O$_{41}$ are staggered
($\delta$=0) as described above.
This simple model also explains the temperature dependence of the
Bragg intensity at (0,1,0), (0,2,2), and (0,0,4).
As shown in Fig. 5, the observed intensity has a finite value at
the temperature where the intensity shows a minimum. The
minimum value of the calculated intensity should become zero as
in Fig. 6b. This is probably due to higher-order neutrons
or a small distortion in the chain. \cite{ta13}
We also calculated the intensity by assuming lattice
distortions which could cause the spin dimerization in the chain.
The intensity calculated at (0,0,2) is decreased when the
lattice distortions are introduced, which is inconsistent with the
experimental results in Sr$_{14}$Cu$_{24}$O$_{41}$.

In summary, we have studied magnetic and structural
properties of the chains in
Sr$_{14-x}$Y$_x$Cu$_{24}$O$_{41}$ (0$\leq x \leq$1). We
observed that when yttrium is substituted for strontium, strong and
sharp magnetic inelastic peaks which originate from the dimerized
state in the chain become broader. The
peaks become broader only in energy but not in momentum space.
This means that the dimerized state becomes unstable but the spin
correlations are unchanged with yttrium substitution. It was also
observed that nuclear Bragg peak intensities originating from the
chains show strong temperature and yttrium concentration
dependence. We proposed a model that the chains shift along the
$c$ axis with temperature and yttrium substitution to explain these
behavior.

\section*{Acknowledgments}
We would like to thank H.-J. Mikeska and Y. Kitaoka for many
helpful discussions.
M. M., S. M. S., and G. S. would like to thank H. Mook, S. Nagler,
and A. Tennant for their warm hospitality during their stay at Oak
Ridge National Laboratory. This work was partially supported
by the U. S.-Japan Cooperative Program on Neutron
Scattering operated by the U. S. Department of
Energy and the Japanese Ministry of Education, Science,
Sports, and Culture, and by the NEDO International Joint
Research Grant. Work at Brookhaven National Laboratory
was carried out under Contract No. DE-AC02-76CH00016,
Division of Material Science, U. S. Department of Energy.
Part of the study was performed at Oak Ridge National Laboratory
which is supported by the Department of Energy, Division of
Materials Sciences under Contract No. DE-AC05-96OR22464.

\begin{figure}
\caption{(a) The chain and the ladder of copper ions
in Sr$_{14}$Cu$_{24}$O$_{41}$, respectively.
Filled circles represent copper atoms and open circles
oxygen atoms. The dashed rectangles represent the
universal unit cell in the (010) crystallographic plane.
Here, $c_{chain}$ and $c_{ladder}$ represent the
lattice constant $c$ for the subcells which contain the
chain and the ladder, respectively.
(b)Temperature dependence of magnetic
susceptibility with the external magnetic field parallel
to the $c$ axis in single crystals of
Sr$_{14-x}$Y$_x$Cu$_{24}$O$_{41}$ ($x$=0, 0.25, and 1.0).}
\label{fig1}
\end{figure}

\begin{figure}
\caption{Inelastic neutron scattering spectra of
Sr$_{14-x}$Y$_x$Cu$_{24}$O$_{41}$ ($x$=0 and 0.25)
at $T$=10 K observed at (0,3,0.085), (0,3,0.17), and (0,3,0.25).
The solid and broken lines are the results of fits to two Gaussians.
The factor for the sample volume, which was estimated
from phonon intensities, is corrected. The inset shows the
measured dispersion (ref. 12) as a function of chain index
$L_{chain}$ in Sr$_{14}$Cu$_{24}$O$_{41}$.}
\label{fig2}
\end{figure}

\begin{figure}
\caption{Inelastic neutron scattering spectra of
Sr$_{13}$Y$_1$Cu$_{24}$O$_{41}$ at $T$=10 K.
A constant-$Q$ scan observed at (0,3,-0.14) (a) and a
constant-$E$ scan at $\Delta E$= 8 meV (b). The solid lines
are guides to the eye.}
\label{fig3}
\end{figure}

\begin{figure}
\caption{Inelastic neutron scattering spectra of
Sr$_{14-x}$Y$_x$Cu$_{24}$O$_{41}$ ($x$=0 and 0.25)
at $T$=10 K observed at (0,0,1.1)$_{ladder}$. The solid and
broken lines are the results of fits to single Gaussian.
The factor for the sample volume is corrected.}
\label{fig4}
\end{figure}

\begin{figure}
\caption{Temperature dependence of the peak intensities measured
at (0,0,2) in Sr$_{14-x}$Y$_x$Cu$_{24}$O$_{41}$
($x$=0, 0.10, and 0.25).}
\label{fig5}
\end{figure}

\begin{figure}
\caption{(a) A proposed model for the structural change in the
chain. Open circles represent the copper ions. Note that the oxygen
ions are not shown here. (b) The calculated intensity of
(0,0,2) as a function of $\delta$.}
\label{fig6}
\end{figure}

\begin{references}
\bibitem[*]{byline}Permanent address: Department of Physics,
Aoyama-Gakuin University, Chitosedai, Setagaya-ku, Tokyo
157, Japan.
\bibitem{ta10}E. M. McCarron, III, M. A. Subramanian,
J. C. Calabrese, and R. L. Harlow, Mat. Res. Bull. {\bf 23},
1355 (1988).
\bibitem{ta13}T. Siegrist, L. F. Schneemeyer, S. A. Sunshine,
J. V. Waszczak, and R. S. Roth, Mat. Res. Bull. {\bf 23}, 1429
(1988).
\bibitem{ta16}M. Azuma, Z. Hiroi, M. Takano, K. Ishida,
and Y. Kitaoka, Phys. Rev. Lett. {\bf 73}, 3463 (1994).
\bibitem{ecc}R. S. Eccleston, M. Azuma, and M. Takano,
Phys. Rev. B. {\bf 53}, R14721 (1996).
\bibitem{kuma}K. Kumagai, S. Tsuji, M. Kato, and Y. Koike,
Phys. Rev. Lett. {\bf 78}, 1992 (1997).
\bibitem{kita}S. Matsumoto, Y. Kitaoka, K. Magishi, K. Ishida,
K. Asayama, M. Uehara, T. Nagata, and J. Akimitsu, (preprint).
\bibitem{ta1}For a review, see E. Daggoto and T. M. Rice,
Science {\bf 271}, 618 (1996).
\bibitem{ta2}T. M. Rice, S. Gopalan, and M. Sigrist, Europhys.
Lett. {\bf 23}, 445 (1993).
\bibitem{ta3}M. Sigrist, T. M. Rice, and F. C. Zheng, Phys.
Rev. B {\bf 49}, 12058 (1994).
\bibitem{aki}M. Uehara, T. Nagata, J. Akimitsu, H. Takahashi,
N. Mori, and K. Kinoshita, J. Phys. Soc. Jpn. {\bf 65}, 2764 (1996).
\bibitem{matsu0}M. Matsuda and K. Katsumata, Phys.
Rev. B {\bf 53}, 12201 (1996).
\bibitem{matsu1}M. Matsuda, K. Katsumata, H. Eisaki,
N. Motoyama, S. Uchida, S. M. Shapiro, and G. Shirane, Phys.
Rev. B {\bf 54}, 12199 (1996).
\bibitem{carter}S. A. Carter, B. Batlogg, R. J. Cava, J. J. Krajewski,
W. F. Peck, Jr., and T. M. Rice, Phys. Rev. Lett. {\bf 77}, 1378
(1996).
\bibitem{matsu2}M. Matsuda, K. Katsumata, T. Yokoo,
S. M. Shapiro, and G. Shirane, Phys. Rev. B {\bf 54}, R15626
(1996).
\bibitem{mike}H.-J. Mikeska and U. Neugebauer,
(to appear in Physica B).
\bibitem{cox}D. E. Cox $et$ $al$., (private communication).
\bibitem{ta25}M. Kato, K. Shiota, and Y. Koike, Physica
{\bf 258C}, 284 (1996).
\end{references}
\end{document}